\documentclass[11pt,twoside]{article}
\usepackage{asp2010}

\aspcpryear{2011}
\aspvoltitle{ADASS XX (to appear)}
\aspvolauthor{Evans, I.~N. and Accomazzi,~A. and Mink, D.~J. and Rots, A.~H.}

\resetcounters

\begin{document}

\title{A Journal for the Astronomical Computing Community?}
\author{Norman Gray$^1$ and Robert G. Mann$^2$}
\affil{$^1$School of Physics and Astronomy, University of Glasgow, UK\\
$^2$Institute for Astronomy, University of Edinburgh,  Edinburgh UK}

\begin{abstract}
One of the Birds of a Feather (BoF) discussion sessions at ADASS XX considered  whether a new journal is needed to serve the astronomical computing community. In this paper we discuss the nature and requirements of that community, outline the analysis that led us to propose this as a topic for a BoF, and review the discussion from the BoF session itself. We also present the results from a survey designed to assess the suitability of astronomical computing papers of different kinds for publication in a range of existing astronomical and scientific computing journals. The discussion in the BoF session was somewhat inconclusive, and it seems likely that this topic will be debated again at a future ADASS or in a similar forum.
\end{abstract}

\section{Introduction} 

The ADASS conference series is approaching the dawn of its third decade in robust health. The past twenty years have seen a marked increase in the importance of computation in support of astronomical research, and this trend seems set to continue, as the data volumes emerging from detectors and simulation codes increase exponentially. 
 The ADASS proceedings volumes provide a very valuable
record of each year's conference, but imperfectly record the activities of
the ADASS community, for a number of reasons: (i) appearing up to a
year after the conference, they often present out-of-date snapshots of rapidly-developing
projects; (ii) being unrefereed, there is no quality threshold, nor
are authors pushed to justify and elaborate where needed to provide
the best account of their material; (iii) being tied to the annual
conference cycle, projects are reported upon when the opportunity
arises, not when they have reached appropriate milestones; and (iv)
having restricted page lengths, topics receive only brief coverage.

This matters for at least two reasons. Firstly, material from most
ADASS conference papers will be published nowhere else, so valuable
technical lessons risk being lost. Secondly, as more people pursue a
career in astronomical computing, it becomes more important that they
have a means of recording their attainment and a track record of
refereed journal papers, with associated citation statistics, is what is most
readily understood by potential employers and assessors of promotion applications. 

Open source publishing systems make the establishment of a
community-driven astronomical computing journal possible, but is it
necessary? Few papers on computational topics appear in mainstream astronomy journals, but is that a reflection of the journals' editorial policies or a lack of interest or confidence on the part of the community? Would mainstream astronomy journals publish more technical
papers if they were submitted? Are there (scientific) computing journals that would
welcome these papers? What are the benefits of refereeing to this
community and would it devote the time needed to referee? Does the
lack of refereed publications hinder the career progress of its
members?

\section{The community}

To the extent that there is an astronomical software community, it is
represented by its attendance at, and support for, ADASS.  Indeed,
the primary publication outlet for many, or
perhaps even most, of the ADASS attendees appears to be the ADASS
proceedings, rather than a cluster of specialised or conventional
journals.  This is an odd situation for an academic discipline, and
so the non-appearance of (the presumed non-null set of) software articles worthy of journal publication
may have several possible explanations.
\begin{enumerate}
\item \label{softwareoutputs}The  primary output from the discipline is software, not
  articles: are these a better (or indeed usable) metric for recognition?
\item \label{service} The members of the community tend to be in
  service roles -- from those performing
  routine software development, to the managers of important parts of the
  astronomical community's infrastructure -- and so their career
  advancement may depend on publication to a lesser extent than conventional astronomers.
\item \label{outlets}There may be no suitable publication outlets, since the existing
  astronomical journals are uninterested in publishing what they
  regard as computing science, and computing science journals are
  uninterested in publishing such applied work.
\item \label{expectations}The community is perhaps such that there is not, or not yet, any
  \emph{expectation} that software results will be published in
  journal form, and the community has therefore not developed any
  shared intuitions about what work is sufficiently valuable, or
  sufficiently interesting, for formal dissemination and careful preservation.
\end{enumerate}
Problem~\ref{softwareoutputs} is too different a question to be considered here.
Problem~\ref{service} may have been true in the past, but it is surely
becoming less true, partly because there is more career crossover between
software and observational astronomy now, than there has been in the
past, and because with astronomy's accelerating move towards HEP-scale
experimentation, and the repeated warnings of the forthcoming `data
deluge', a broader range of software technologies have become integral
to present and future astronomical practice.  What this means in turn is that
there is a growing number of individuals whose principal intellectual
excitement, and whose principal contribution to astronomy, is
\emph{via} innovative software and system development.  These
people are not observers or theorists, nor are they computer
scientists, but are instead something in between.  The term
\emph{astroinformatics} seems convenient.

\section{The survey}
\label{s:survey}

To test hypothesis~\ref{outlets},
we mailed the editors of MNRAS, A\&A, ApJ, Earth Science Informatics,
Experimental Astronomy, CODATA Data Science Journal, Astronomische
Nachrichten and PASP, with five titles and abstracts from last year's
ADASS proceedings.  This set of five articles was chosen because each
seemed typical of one or other class of publication commonly
presented at ADASS, and we asked the editors to assess whether the
subject matter of the article, independently of its body, would be
deemed sufficiently in scope for it to be passed on to a referee.

The article topics were:
 {\bf alg} -- software implementation of scientific algorithms (Bayesian techniques for classification);
 {\bf app} -- application progress report (WWT update);
 {\bf pipe} -- pipeline features and recent developments (detailed report of new IDL pipeline features);
 {\bf gen} -- application of general computing technologies to astronomy (application of Java and HPC techniques to a specific mission);
 {\bf inf} -- development and use of astronomy-specific `infrastructure' (benefits of HEALPix in a particular application).

\begin{table}[!ht]
\caption{Summary of journal survey results}

\begin{center}
{\small
\begin{tabular}{cccccc}
\tableline
\noalign{\smallskip}
& alg & app & pipe & gen & inf\\
\tableline
\noalign{\smallskip}
A\&A$^1$&yes?&no?$^2$&?&no&?\\
\noalign{\smallskip}
MNRAS$^3$&no&no&no&no&no\\
\noalign{\smallskip}
ApJ$^4$&yes?&no?$^5$&yes??&no&yes??\\
\noalign{\smallskip}
ESIn$^6$&yes&yes&no?&yes&yes\\
\noalign{\smallskip}

DSJ$^7$&yes&yes&no?&yes&yes\\
\noalign{\smallskip}

PASP$^8$&yes&yes&no?&yes&yes\\
\noalign{\smallskip}
\tableline
\end{tabular}
}
\end{center}
\end{table}

 The
responses are summarised in Table 1, and are published in full at the following website: {\tt http://www.roe.ac.uk/$\sim$rgm/bof.html}.

Notes to Table 1:
1.~Could appear in `Astronomy Instrumentation' section; must be ``of interest to a
sizable fraction of the A\&A audience'';
2.~Issue of VO tools to be discussed soon by editors;
3.~``Descriptions of new software appear only if accompanied by new science derived using it''; 
4.~``[O]ur enthusiasm for techniques papers tend to fall off as they become less
concerned with direct results and more divorced from ongoing science
projects'', but OK if ``the paper will be interesting even if the
particular instrument never actually gets built'';
5.~Might be considered for a WWT special issue;
6.~``The topics \dots easily fall under the focus of
this journal. [\dots] We consider astronomy informatics a sister domain,
related to Earth Science Informatics''; 
7.~``[D]efinitely in the scope'' of DSJ;
8.~``We have indeed
accepted articles such as the 5 that you sent''.  No replies were
received from Experimental Astronomy or from Astronomische Nachrichten.

\section{Summary of BoF discussion}
Opinion varied widely as to the necessity of a new journal, but there was general 
enthusiasm at the idea of participating -- as readers, authors, referees, and, 
in some cases, as editors -- in such a new journal, if one were to exist. 

Some people felt that the increasing importance of computational techniques in 
astronomy today necessitates the creation of a dedicated journal, and the
analogy was made with particle physics, which has long since made a definite split 
between journals for science results and journals for technical material (experimental 
details, as well as analysis software, etc). The latter stream is highly valued, both 
for providing a means of sharing and recording technical knowledge, and for rewarding 
the efforts of more technical staff, prompting the suggestion that astronomy would benefit 
from the same system.
The case was made that previous attempts to provide such outlets have not met with
great enthusiasm from the ADASS community. The AIP's ``Computers in Physics'' 
journal had a similar intention to that proposed here (although covering all of physics,
rather than just astronomy), but few astronomy-related papers were published there, 
and the same goes now for Experimental Astronomy. 

There was general agreement that the ADASS community would benefit from producing 
more refereed papers, but most people felt that the opportunities provided by existing
journals should be exhausted before serious consideration is made of starting a new
one. This might even by coordinated -- whether through the dedication of special issues,
or more informally amongst authors -- in an attempt to produce a journal with a critical
mass of material about astronomy computing. It is not clear whether the increasing 
importance of this domain is better highlighted by the creation of a dedicated journal, 
or by making a significant presence within an existing journal.  

Many people see PASP as the most appropriate outlet for papers on topics requiring
a fuller treatment than allowed by the ADASS proceedings, and others mentioned that 
papers on algorithms can find a home in mainstream astronomy journals, so the problem of
\emph{excluded} material is largely restricted to descriptions of pipeline software and the like, whose 
details should be recorded and made available to their users, but which may lack the
conceptual novelty required by most journals
(that is, problem~\ref{outlets} may apply only to articles of type \textbf{pipe} in Table~1).
This both records and advertises the authoring software group's contributions to the astronomical enterprise.

A requirement was identified for an outlet for publishing lessons learnt of the ``we did
this, but it didn't work because of these reasons'' sort. That could be provided by a
non-refereed section of a new journal, or, equally, by postings to astro-ph. There is
already an ``instruments and methods'' chapter there, that is currently poorly used by the
ADASS community, but it could be transformed by greater use into a suitable vehicle for
knowledge exchange within the community. 

Another advantage of publishing in astro-ph is that research astronomers are used to
looking there, which would be an advantage for some papers in this domain, although,
equally, some others may benefit more from publication in a journal (such as Earth
Sciences Informatics or CODATA Data Science Journal) that is read by people working on
analogous topics in related domains.

\section{Conclusions}

Our survey reveals that, although the high-impact `big three'
journals do not see ADASS material as naturally in their scope,
there are other journals which would be perfectly willing to consider
articles, and that conclusion was shared by the BoF participants.  There seems little present need for a new journal.

The key question then becomes number~\ref{expectations}: why does the community not publish in the
journals that are available to it? We look forward to a spike in
astroinformatics journal articles in 2011.

\acknowledgements
We are grateful to the journal editors who were generous in
examining the sample papers we sent them, and in clarifying the goals
they have for their journals.
We are grateful to the many people who attended the BoF in Boston, and
for their thoughtful comments.  
We are particularly grateful to Rob Seaman for his persistence in
unearthing more journals than we believed possible.

\bibliographystyle{asp2010}

\end{document}